\documentclass[preprint,showkeys,lengthcheck,nofootinbib,twocolumn,notitlepage,floatfix,superscriptaddress]{revtex4-1}
\usepackage{graphics,graphicx}
\usepackage{amsmath, amssymb}
\usepackage{multirow}
\usepackage{color}
\usepackage{verbatim}
\usepackage{hyperref}
\usepackage[normalem]{ulem}  
\usepackage{ulem}
\usepackage{color}
\usepackage{cancel}
\usepackage{mathtools}
\usepackage{epstopdf}

\renewcommand\sout{\bgroup\color{blue} \ULdepth=-.5ex \ULset}
\def\slashchar#1{\setbox0=\hbox{$#1$}  
\dimen0=\wd0     
\setbox1=\hbox{/} \dimen1=\wd1  
\ifdim\dimen0>\dimen1   
\rlap{\hbox to \dimen0{\hfil/\hfil}} 
#1     
\else     
\rlap{\hbox to \dimen1{\hfil$#1$\hfil}} 
/      
\fi}

\newcommand{\dd}{\mathrm{d}}

\begin{document}

\title{Chiral symmetry restoration and the \texorpdfstring{$\ensuremath \Delta$}{} matter formation in neutron stars}

\date{\today}
\author{Micha\l{} Marczenko}
\email{michal.marczenko@uwr.edu.pl}
\address{Incubator of Scientific Excellence - Centre for Simulations of Superdense Fluids, University of Wroc\l{}aw, plac Maksa Borna 9, PL-50204 Wroc\l{}aw, Poland}
\author{Krzysztof Redlich}
\address{Institute of Theoretical Physics, University of Wroc\l{}aw, plac Maksa Borna 9, PL-50204 Wroc\l{}aw, Poland}
\author{Chihiro Sasaki}
\address{Institute of Theoretical Physics, University of Wroc\l{}aw, plac Maksa Borna 9, PL-50204 Wroc\l{}aw, Poland}

\begin{abstract}
	We analyze the effects of chiral symmetry restoration in hadronic matter, including the lowest-lying baryonic resonance $\Delta$ based on the parity doublet model. We study the role of $\Delta$ and its chiral partner on the equation of state (EoS) of dense matter under neutron star (NS) conditions of $\beta$-equilibrium and charge neutrality. We find that the softening of the EoS driven by the early onset of $\Delta$ matter due to partial restoration of chiral symmetry allows accommodating the modern multi-messenger astrophysical constraints on the mass, radius, and tidal deformability. The softening above the saturation density is accompanied by subsequent stiffening at high densities. We also find that the matter composition in the NS cores may be different upon variations of the repulsive interactions of $\Delta$ baryons in hadronic matter.
\end{abstract}
\maketitle 

\section{Introduction}
\label{sec:introduction}

Neutron stars (NSs) are unique extraterrestrial laboratories to probe matter under extreme conditions. Recently, the possibility of the formation of baryons other than nucleons in the cores of NSs has become one of the central issues in modern nuclear astrophysics~\cite{Oertel:2016bki}. Many extensive studies address the appearance of hyperons, especially in the context of the notorious hyperon puzzle~(see, e.g.,~\cite{Glendenning:1984jr, Drago:2013fsa, Baym:2017whm, Lattimer:2006qiu, Raduta:2017wpp, Fortin:2017cvt, Oertel:2014qza, Gomes:2014aka, Katayama:2015dga, Bombaci:2016xzl}). In contrast, much less research has been conducted on the influence of $\Delta(1232)$ resonance on the gross properties of the equation of state (EoS). In fact, for many years, its possible contribution has been rather ignored because early results suggested that $\Delta(1232)$ resonance appears at densities beyond the typical central densities of compact objects~\cite{Glendenning:1984jr}. However, recent studies based on the relativistic mean-field (RMF) model~\cite{Drago:2014oja} and microscopic approaches~\cite{Li:2019tjx} suggest an early onset of $\Delta(1232)$ resonance, which, similarly to hyperons, leads to a significant softening of the EoS. In the last decade, a number of studies of $\Delta(1232)$ in dense matter were conducted within various approaches~\cite{Drago:2014oja, Li:2018qaw, Motta:2019ywl, Li:2020ias, Li:2019tjx, Cai:2015hya, Zhu:2016mtc, Sahoo:2018xeu, Sen:2021bms, Malfatti:2020onm, Ribes:2019kno, Schurhoff:2010ph, Drago:2013fsa, Marczenko:2021uaj}. The consequences of the appearance of $\Delta(1232)$ resonance are found to be crucial for the properties of the EoS and the structure of NSs. Similar to the hyperon contribution, the early appearance of $\Delta(1232)$ resonance creates a tension with the maximum NS mass. Consequently, obtained maximal masses are far below the mass of the PSR J0740+6620 pulsar~\cite{Demorest:2010bx, Antoniadis:2013pzd, Fonseca:2016tux, Cromartie:2019kug, Fonseca:2021wxt}. On the other hand, the radius and tidal deformability of a $1.4~M_\odot$ NS can be significantly reduced due to the softening of the EoS linked to the early appearance of $\Delta(1232)$ resonance at low to intermediate densities~\cite{Li:2018qaw, Li:2019tjx}. 

The recent advancements of multi-messenger astronomy on different dense-matter astrophysical sources have led to further improvements in constraining the EoS at low temperature and high density. The modern observatories for measuring masses and radii of compact objects, the gravitational wave interferometers of the LIGO/Virgo Collaboration (LVC)~\cite{Abbott:2018exr, LIGOScientific:2018hze}, and the X-ray observatory Neutron star Interior Composition Explorer (NICER) provide new powerful constraints on the neutron-star mass-radius profile~\cite{Riley:2019yda, Miller:2019cac, Miller:2021qha, Riley:2021pdl}. These stringent constraints allow for a more systematic study of the influence of the formation of various degrees of freedom inside the cores of NSs. Therefore, it is crucial to explore the underlying role of $\Delta(1232)$ resonance in dense matter under extreme conditions.

Central densities of neutron stars lie up to a few times of normal nuclear density so that it is to be expected that baryons, in particular nucleons and $\Delta$, change their properties due to the restoration of chiral symmetry. The recent lattice QCD (LQCD) results~\cite{Aarts:2015mma, Aarts:2017rrl, Aarts:2018glk} exhibit a clear manifestation of the parity doubling structure for the low-lying baryons around the chiral crossover. The observed behavior of parity partners is likely an imprint of the chiral symmetry restoration in the baryonic sector of QCD and is expected to occur also in cold dense matter, including neutron-star conditions. Such properties of the chiral partners can be described in the framework of the parity doublet model~\cite{Detar:1988kn, Jido:1999hd, Jido:2001nt}. The model has been applied to hot and dense hadronic matter, neutron stars, as well as the vacuum phenomenology of QCD~\cite{Dexheimer:2007tn, Gallas:2009qp, Paeng:2011hy, Sasaki:2011ff, Gallas:2011qp, Zschiesche:2006zj, Benic:2015pia, Marczenko:2017huu, Marczenko:2018jui, Marczenko:2019trv, Marczenko:2020wlc, Marczenko:2020jma, Marczenko:2020omo, Marczenko:2021uaj, Mukherjee:2017jzi, Mukherjee:2016nhb, Dexheimer:2012eu, Steinheimer:2011ea, Weyrich:2015hha, Sasaki:2010bp, Yamazaki:2018stk, Yamazaki:2019tuo, Ishikawa:2018yey, Steinheimer:2010ib, Giacosa:2011qd, Motohiro:2015taa, Minamikawa:2020jfj, Minamikawa:2021fln}.

In this work, we explore the implications of dynamical restoration of chiral symmetry within the hadronic phase. To this end, we employ the parity doublet model for the nucleon and $\Delta$~\cite{Jido:1999hd, Takeda:2017mrm} and explore their implications on the structure of neutron stars. We demonstrate how modern constraints on the masses, radii, and tidal deformability of neutron stars can be achieved in this framework.

This paper is organized as follows. In Sec.~\ref{sec:pd_model}, we introduce the parity doublet model for nucleon and $\Delta(1232)$ resonance. In Sec.~\ref{sec:eos}, we discuss the obtained numerical results on the equation of state and chiral structure in the isospin-symmetric matter and under neutron-star conditions. In Sec.~\ref{sec:neutron_stars}, we discuss the obtained neutron-star relations and confront the results with recent observations. Finally, Sec.~\ref{sec:summary} is devoted to summary and conclusions.

\section{Parity doublet model}
\label{sec:pd_model}

\begin{table*}[t!]\begin{center}\begin{tabular}{|c|c|c|c|c|c|c|c|}
  \hline
  $m_N~$[MeV] & $m_N^\star~$[MeV] & $m_\Delta~$[MeV] & $m_\Delta^\star~$[MeV] & $m_\pi~$[MeV]  & $m_\omega~$[MeV] & $m_\rho~$[MeV] & $f_\pi~$[MeV] \\ \hline\hline
  939   & 1500  & 1232 & 1700 & 140       & 783        & 775 & 93  \\ \hline
  \end{tabular}\end{center}
  \caption{Physical vacuum inputs and the parity doublet model parameters used in this work.}
  \label{tab:vacuum_params}
\end{table*}

\begin{table*}[t!]\begin{center}\begin{tabular}{|c|c|c|c|}
  \hline
  $n_0~[\rm fm^{-3}]$ & $E/A- m_N~[\rm MeV]$ & $K~$[MeV] & $E_{\rm sym}~$[MeV] \\ \hline\hline
  0.16                   & -16                    & 240       & 31 \\ \hline
  \end{tabular}\end{center}
  \caption{Properties of the nuclear ground state at $\mu_B=923~$MeV and $\mu_Q=0$, and the symmetry energy used in this work.}
  \label{tab:vacuum_params2}
\end{table*}

In this section, we briefly introduce the parity doublet model for nucleon and $\Delta(1232)$ resonance capable of describing the chiral symmetry restoration, following Ref.~\cite{Takeda:2017mrm}. The model is composed of the baryonic parity doublets for nucleon and $\Delta(1232)$ resonance, and mesons as in the Walecka model~\cite{Walecka:1974qa}. The spontaneous chiral symmetry breaking yields the mass splitting between the two baryonic parity partners, in each parity doublet with given spin. In this work, we consider a system with $N_f=2$. The baryonic degrees of freedom are coupled to the chiral fields $\left(\sigma, \boldsymbol\pi\right)$, the vector-isoscalar field ($\omega_\mu$), and the vector-isovector field ($\boldsymbol \rho_\mu$). The thermodynamic potential of the model in the mean-field approximation reads~\cite{Takeda:2017mrm}
\begin{equation}\label{eq:thermo_potential}
	\Omega = V_\sigma + V_\omega + V_\rho + \sum_{x=N,\Delta}\Omega_x\rm,
\end{equation}
with the index $N$ labeling collectively positive-parity and negative-parity spin-$1/2$ nucleons, i.e., $N\in \lbrace p,n;p^\star,n^\star \rbrace$, and spin-$3/2$ $\Delta$'s, i.e., \mbox{$\Delta \in \lbrace\Delta_{++,+,0,-};\Delta^\star_{++,+,0,-}\rbrace$}. The negative-parity states are marked with asterisks. The mean-field potentials in Eq.~\eqref{eq:thermo_potential} read
\begin{subequations}\label{eq:potentials}
\begin{align}
  V_\sigma &= -\frac{\lambda_2}{2}\sigma^2 + \frac{\lambda_4}{4}\sigma^4 - \frac{\lambda_6}{6}\sigma^6 - \epsilon\sigma \textrm,\label{eq:potentials_sigma}\\
  V_\omega &= -\frac{m_\omega^2 }{2}\omega^2\textrm,\label{eq:potentials_omega}\\
  V_\rho &= - \frac{m_\rho^2}{2}\rho^2\textrm,\label{eq:potentials_rho}
\end{align}
\end{subequations}
where $\lambda_2 = \lambda_4f_\pi^2 - \lambda_6f_\pi^4 - m_\pi^2$, and $\epsilon = m_\pi^2 f_\pi$. $m_\pi$, $m_\omega$, and $m_\rho$ are the $\pi$, $\omega$, and $\rho$ meson masses, respectively, and $f_\pi$ is  the pion decay constant. The parameters $\lambda_4$ and $\lambda_6$ are fixed by the properties of the nuclear ground state. We note that the six-point scalar interaction term in Eq.~\eqref{eq:potentials_sigma} is essential in order to reproduce the experimental value of the compressibility \mbox{$K=240\pm20~$MeV~\cite{Shlomo, Motohiro:2015taa}}.  In Eqs.~\eqref{eq:potentials_omega} and \eqref{eq:potentials_rho}, $\omega$ and $\rho$ are the only non-vanishing expectation values of the $\omega_\mu$ and $\boldsymbol \rho_\mu$ in the mean-field approximation, respectively. The kinetic part of the thermodynamic potential, $\Omega_x$, reads
\begin{equation}\label{eq:thermokin}
\Omega_x = \gamma_x \int \frac{\dd^3p }{(2\pi)^3} T \left(\ln\left(1-f_x\right) + \ln \left(1 - \bar f_x\right)\right)\rm,
\end{equation}
where the factors $\gamma_N = 2$ and $\gamma_\Delta=4$ denote the spin degeneracy of both parity partners for nucleons and $\Delta$'s, respectively. The particle (antiparticle) Fermi-Dirac distribution function reads
\begin{subequations}
\begin{align}
	f_x &= \frac{1}{1+e^{\beta(E_x-\mu_x)}}\rm ,\\
	\bar f_x &= \frac{1}{1+e^{\beta(E_x+\mu_x)}}\rm,
\end{align}
\end{subequations}
with $\beta$ being the inverse temperature, the dispersion relation $E_x = \sqrt{\boldsymbol p^2+m_x^2}$ and $\mu_x$ is the effective chemical potential.

The masses of the positive- and negative-parity chiral partners are given by
\begin{equation}\label{eq:doublet_mass}
	m^x_\pm = \frac{1}{2}\left[\sqrt{\left(g_1^x+g_2^x\right)^2\sigma^2 + 4\left(m_0^x\right)^2} \mp \left(g^x_1-g^x_2\right)\sigma\right] \textrm,
\end{equation}
where $\pm$ sign denotes parity and $x=N,\Delta$ for nucleons and $\Delta$, respectively. The positive-parity nucleons are identified as the positively charged and neutral $N(938)$ states: proton ($p$) and neutron ($n$). Their negative-parity counterparts, denoted as $p^\star$ and $n^\star$, are identified as $N(1535)$ resonance~\cite{ParticleDataGroup:2020ssz}. The positive-parity $\Delta$ states are identified with $\Delta(1232)$. Their negative-parity chiral partners, $\Delta^\star$, are identified with $\Delta(1700)$~\cite{ParticleDataGroup:2020ssz}. For given chirally invariant mass, $m_0^x$, the parameters $g_1^x$ and $g_2^x$ are determined by the corresponding vacuum masses (see Table~\ref{tab:vacuum_params}).

The effective chemical potentials for nucleons and their chiral partners are given by
\begin{subequations}\label{eq:eff_chem_pot_N}
\begin{align}
	\mu_{p} &= \mu_{p^\star} = \mu_B + \mu_Q - g_\omega^N\omega - g_\rho^N\rho\rm,\\
	\mu_{n} &= \mu_{n^\star} = \mu_B - g_\omega^N\omega + g_\rho^N\rho\rm.
\end{align}
\end{subequations}
The effective chemical potentials for $\Delta$ and their chiral partners are given by
\begin{subequations}\label{eq:eff_chem_pot_D}
\begin{align}
	\mu_{\Delta_{++}} &= \mu_{\Delta^\star_{++}} = \mu_B + 2\mu_Q - g_\omega^\Delta\omega - 3g_\rho^\Delta\rho \rm,\\
	\mu_{\Delta_{+}} &= \mu_{\Delta^\star_{+}}  = \mu_B + \mu_Q - g_\omega^\Delta\omega - g_\rho^\Delta\rho\rm,\\
	\mu_{\Delta_{0}} &= \mu_{\Delta^\star_{0}}  = \mu_B - g_\omega^\Delta\omega + g_\rho^\Delta\rho\rm,\\
	\mu_{\Delta_{-}} &= \mu_{\Delta^\star_{-}}  = \mu_B -\mu_Q - g_\omega^\Delta\omega + 3g_\rho^\Delta\rho\rm.
\end{align}
\end{subequations}
The parameters, $g_\omega^x$ and $g_\rho^x$ control the coupling strengths of baryons to $\omega$ and $\rho$ mesons, respectively~\cite{Takeda:2017mrm}.

In-medium profiles of the mean fields are obtained by extremizing the thermodynamic potential in Eq.~\eqref{eq:thermo_potential}, leading to the following gap equations
\begin{subequations}
\begin{align}\label{eq:gap_eqs}
	\frac{\partial \Omega}{\partial \sigma} &= \frac{\partial V_\sigma}{\partial \sigma} + \sum_{x=N,\Delta}s_x \frac{\partial m_x}{\partial \sigma}\textrm,\\
	\frac{\partial \Omega}{\partial \omega} &= \frac{\partial V_\omega}{\partial \omega} + \sum_{x=N,\Delta}g^x_\omega n_x \textrm,\\
	\frac{\partial \Omega}{\partial \rho}   &= \frac{\partial V_\rho}{\partial \rho} + g^N_\rho \left(n_p - n_n + n_{p^\star} - n_{n^\star} \right) \\
	&+g_\rho^\Delta(3n_{\Delta_{++}}+n_{\Delta_+} -n_{\Delta_0} - 3n_{\Delta_-}\nonumber\\
	&\phantom{\;\;\;\;\;\;}+3n_{\Delta^\star_{++}}+n_{\Delta^\star_+} -n_{\Delta^\star_0} - 3n_{\Delta^\star_-})\rm,\nonumber
\end{align}
\end{subequations}
where the scalar and vector densities are
\begin{equation}
	s_x = \gamma_x \int \frac{\mathrm{d}^3 p}{\left(2\pi\right)^3} \frac{m_x}{E_x}\left(f_x + \bar f_x\right)
\end{equation}
and
\begin{equation}\label{eq:vec_dens}
	n_x = \gamma_x \int \frac{\mathrm{d}^3 p}{\left(2\pi\right)^3}\left(f_x - \bar f_x\right) \textrm,
\end{equation}
respectively.

In the grand canonical ensemble, the~thermodynamic pressure is obtained from the thermodynamic potential as \mbox{$P = -\Omega + \Omega_0$}, where $\Omega_0$ is the value of the thermodynamic potential in the vacuum.

The net-baryon number density and net-charge densities for a species $x$ are defined as
\begin{subequations}
\begin{align}
  n^x_B &= -\frac{\partial \Omega_x(T, \mu_B, \mu_Q)}{\partial \mu_B} \textrm,\\
  n^x_Q &= -\frac{\partial \Omega_x(T, \mu_B, \mu_Q)}{\partial \mu_Q} \textrm,
\end{align}
\end{subequations}
respectively, where $\Omega_x$ is the kinetic term in Eq.~\eqref{eq:thermokin}. The particle-density fractions are defined as
\begin{equation}\label{eq:fractions}
  Y_x = \frac{n_B^x}{n_B} \textrm.
\end{equation}
The compressibility, symmetry energy, and its slope at saturation density are given as
\begin{subequations}
\begin{align}
	K &= 9n_0 \frac{\partial \mu_B}{\partial n_B}\Bigg|_{n_B=n_0}\textrm,\\
	E_{\rm sym} &= \frac{1}{2}\frac{\partial^2 \left(\epsilon / n_B \right)}{\partial \delta^2}\Bigg|_{\delta = 0} \textrm,\label{eq:e_sym}\\
  L &= 3 n_0 \frac{\partial E_{\rm sym}}{\partial n_B} \Bigg|_{n_B = n_0}\textrm,
 \end{align}
\end{subequations}
respectively. In Eq.~\eqref{eq:e_sym}, $\epsilon$ is the energy density and the isospin asymmetry parameter $\delta = \sum_i I_i n_i / n_B$, where $I_i$ is the third component of the isospin operator of the i'th species\footnote{We note that this generalized definition reduces to the known definition of the asymmetry parameter, \mbox{$\delta = (n_B^p - n_B^n) / n_B$}, where the negative parity states, $N^\star$ and $\Delta$ states are not populated, and their densities vanish, which is the case at the saturation density.}. We remark that the obtained values of $L_{\rm sym}\approx82~$MeV at saturation for $m_0^N = 550-700~$MeV agree with the commonly considered range of the parameter~\citep{Oertel:2016bki} and are also found in other parity-doublet models~\citep{Motohiro:2015taa}. The most updated estimate, $L_{\rm sym} = 53^{+14}_{-15}~$MeV, is based on combined astrophysical data, PREX-II, and recent effective chiral field theory results~\citep{Essick:2021kjb}. Furthermore, the recent analysis within density functional theory yields $L_{\rm sym} = 54\pm8$~MeV~\citep{Reinhard:2021utv}. We note that the value of the parameter $L_{\rm sym}$ can be decreased, e.g., by introducing an $\omega-\rho$ interaction term to the Lagrangian~\cite{Dexheimer:2018dhb}.

The strength of the $g^N_\omega$ coupling is fixed by the nuclear saturation properties: the saturation density, $n_0$, and the binding energy, $\epsilon/n_B-m_N$; the value of $g^N_\rho$ can be fixed by fitting the value of symmetry energy. The properties of the nuclear ground state and the symmetry energy are compiled in Table~\ref{tab:vacuum_params2}. The couplings of the $\Delta$ resonance to the meson fields are poorly constrained due to limited knowledge from experimental observations. The most advocated constraint was obtained by an analysis of electromagnetic excitations of the $\Delta$ baryon in the framework of relativistic mean-field model~\cite{Wehrberger:1989cd}. It puts a constraint on the relative strength of the scalar and vector couplings. Other phenomenological studies indicate an attractive $\Delta-N$ potential with no consensus on its actual size~\cite{Horikawa:1980cv, OConnell:1990njm, Lehr:1999zr, Nakamura:2009iq}. We note that in the parity doublet model the values of the $\Delta-\sigma$ couplings, $g_1^\Delta$ and $g_2^\Delta$, are uniquely fixed by requiring the vacuum masses of $\Delta$ and $\Delta^\star$. On the other hand, the nature of the repulsive interaction among $\Delta$ isobars and their coupling to the $\omega$ and $\rho$ mean fields are still far from consensus. 

\begin{figure}[t!]
  \centering
  \includegraphics[width=\linewidth]{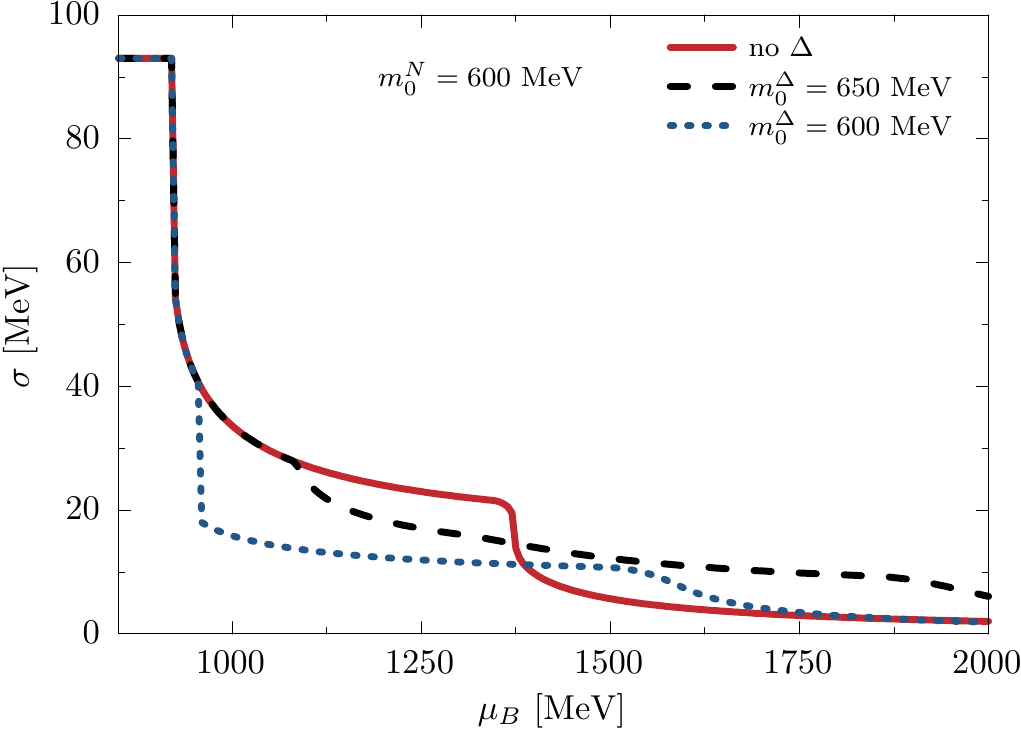}
  \caption{The $\sigma$ mean field in the parity doublet model for $m_0^N=600~$MeV in isospin-symmetric matter at zero temperature as a function of the baryon chemical potential.}
  \label{fig:sigma_600}
\end{figure}

\begin{figure}[t!]
  \centering
  \includegraphics[width=\linewidth]{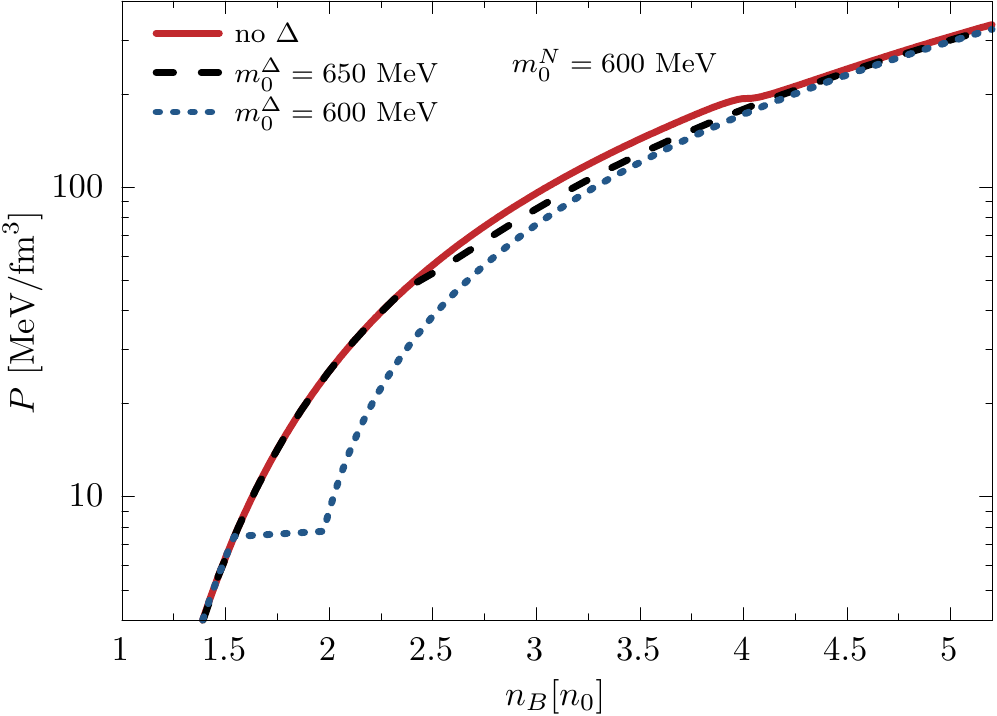}
  \caption{Thermodynamic pressure in isospin-symmetric matter as a function of the net-baryon number density, in units of the saturation density, $n_0=0.16~\rm fm^{-3}$.}
  \label{fig:p_rho_600}
\end{figure}

It is customary to parametrize the $\Delta$-meson couplings in terms of the nucleon-meson couplings:
\begin{subequations}\label{eq:r_delta}
\begin{align}
    g_\omega^\Delta &= R_\Delta g_\omega^N \textrm, \\
    g_\rho^\Delta &= R_\Delta g_\rho^N \textrm.
\end{align}
\end{subequations}
For simplicity, in the present study, we fix $R_\Delta=1$. Detailed discussion of the dependence of $R_\Delta$ is presented in Sec.~\ref{sec:repulsion_delta}. In general, additional repulsion between $\Delta$'s would systematically shift their onset in the stellar sequence to higher densities. This eventually would prevent the neutron stars with $\Delta$ matter from existence in the gravitationally stable branch of the sequence. We note that this effect is similar to the case of repulsive interactions between quarks~\cite{Marczenko:2020jma}.
 
\begin{figure*}[t!]
  \centering
  \includegraphics[width=.475\linewidth]{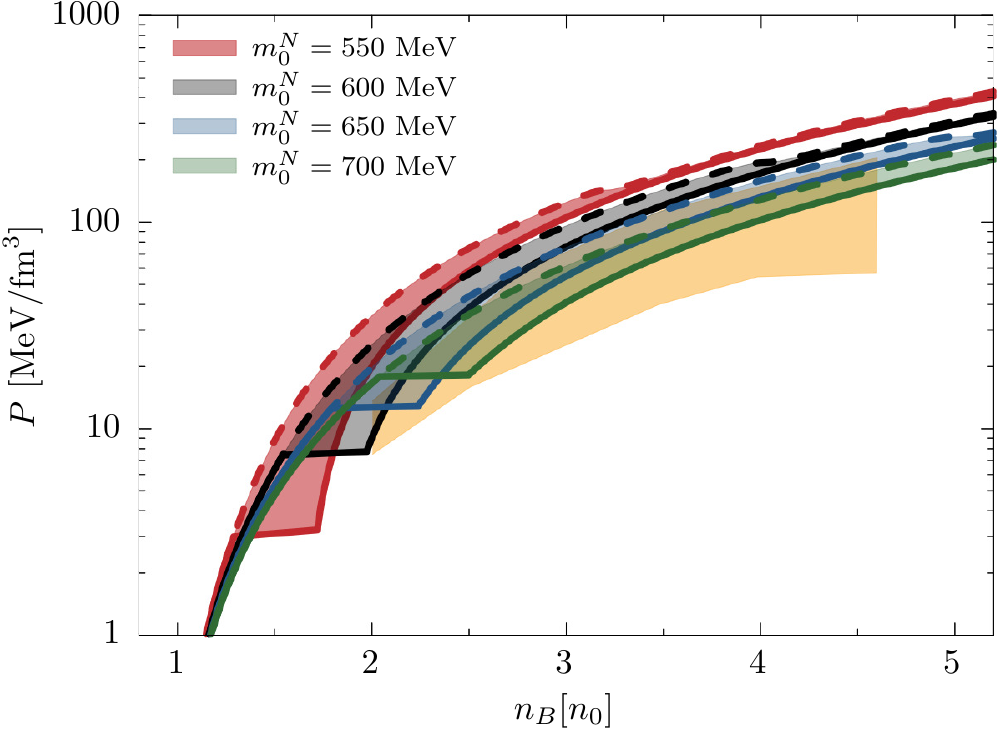}\;\;\;\;
  \includegraphics[width=.475\linewidth]{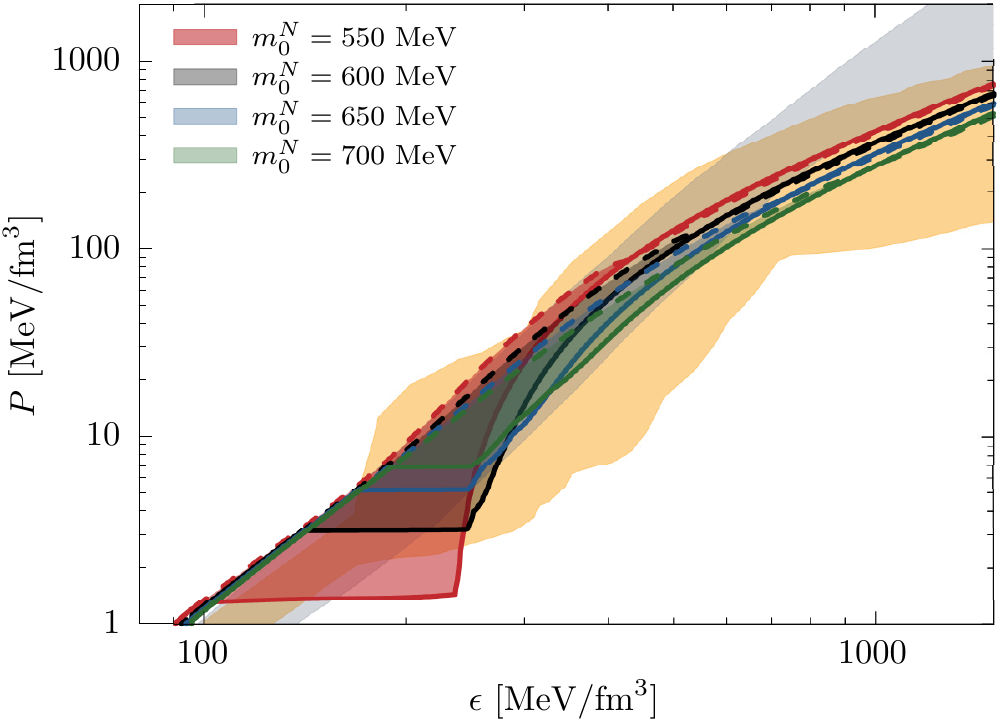}
  \caption{Thermodynamic pressure for isospin-symmetric matter as a function of the net-baryon number density, in units of the saturation density, $n_0=0.16~\rm fm^{-3}$ (left panel), and under the NS conditions of $\beta$-equilibrium and charge neutrality, as a function of the energy density, $\epsilon$, at $T=0$ (right panel). In the left panel, the orange-shaded region shows the flow constraint~\cite{Danielewicz:2002pu}. In the right panel, the orange- and gray-shaded regions show the constraints obtained by~\cite{Annala:2019puf} and~\cite{Abbott:2018exr}, respectively.}
  \label{fig:pressure}
\end{figure*}

In the present work, we take four representative values of $m_0^N=550$, $600$, $650$, $700~$MeV. Because the onset of $\Delta$ matter depends on the value of the chirally invariant mass $m_0^\Delta$~\cite{Takeda:2017mrm}, we systematically study the influence of $\Delta$ on the EoS and compliance with terrestrial constraints from heavy-ion collisions~\cite{Danielewicz:2002pu}, and the astrophysical constraints, i.e., the $2~M_\odot$ and the tidal deformability constraints~\cite{Hebeler:2013nza, Cromartie:2019kug, Fonseca:2021wxt, Abbott:2018exr}. In this work, we put an additional constraint on the chirally invariant mass of $\Delta$. Namely, we require that $m_0^N \leq m_0^\Delta$. Too low values of $m_0^\Delta$ lead to the onset of $\Delta$ matter at subsaturation densities; thus, it spoils the properties of the ground state~\cite{Takeda:2017mrm}.  Setting $m_0^\Delta = \infty$ suppresses the $\Delta$ states and the EoS effectively corresponds to the purely nucleonic EoS. We note that by  the assumption, $L_{\rm sym}$ does not depend on the choice of $m_0^\Delta$.
 
The physical inputs used in this work are summarized in Tabs.~\ref{tab:vacuum_params} and~\ref{tab:vacuum_params2}.

\section{Chiral Structure and Equation of State}
\label{sec:eos}

\begin{figure*}[t!] \centering
    \includegraphics[width=.49\linewidth]{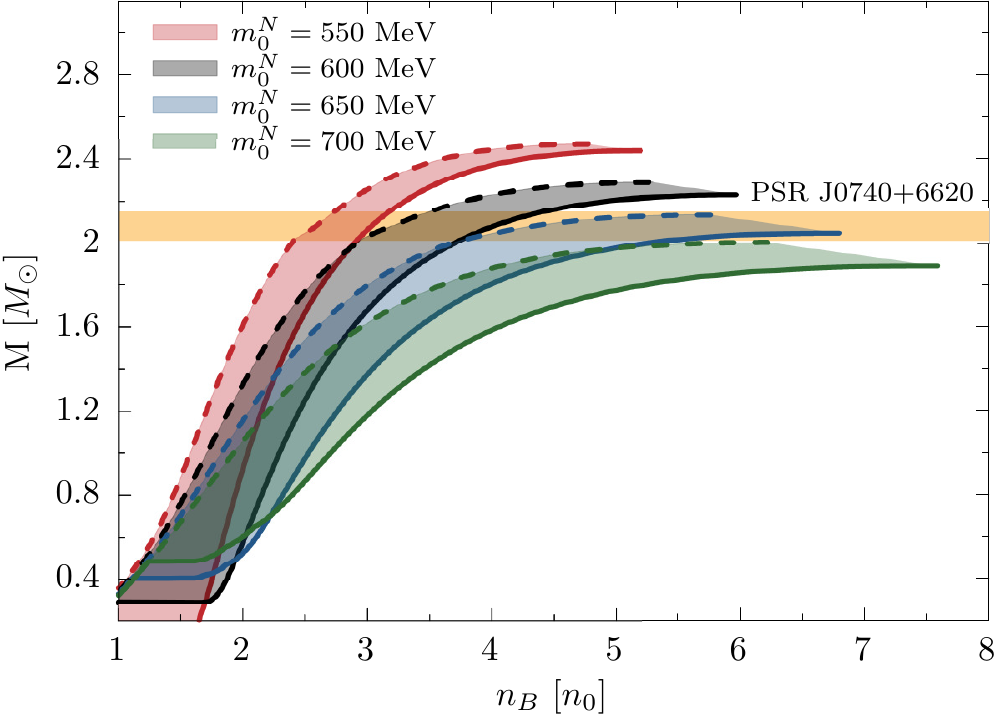}
    \includegraphics[width=.49\linewidth]{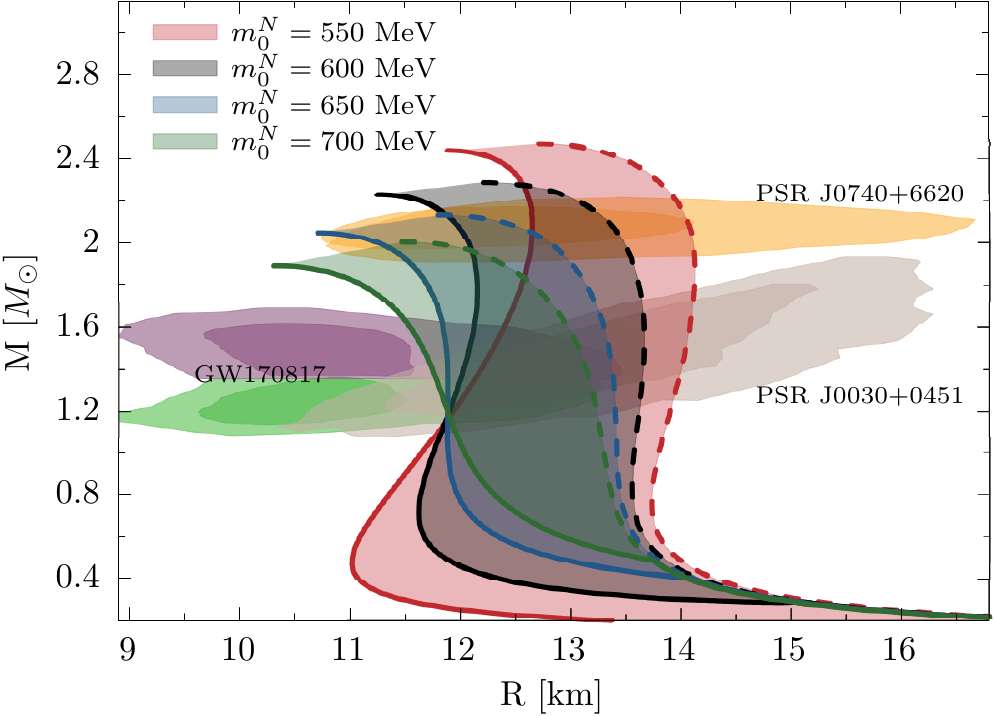}
    \caption{Sequences of masses for neutron stars vs. central net-baryon density (left panel) and radius (right panel) as solutions of the TOV equations. The dashed lines correspond to the purely nucleonic EoSs. The solid lines correspond to the case $m_0^N=m_0^\Delta$. The region spanned between the two lines mark the results obtained for $m_0^N < m_0^\Delta$ in each case. The mass-density and mass-radius curves are plotted up to the maximal mass solutions for given $m_0^N$. The orange band in the left panel shows the PSR J0740+6620 mass constraint $M_{\rm max} = 2.08\pm0.07$~\cite{Fonseca:2021wxt}. The inner (outer) orange band in the right panel shows the $1\sigma$ credibility regions from the NICER analysis of observations of the massive pulsar PSR J0740+6620 as dark orange~\cite{Riley:2021pdl} and light orange~\cite{Miller:2021qha} regions. The inner (outer) green and purple bands show 50\% (90\%) credibility regions obtained from the recent GW170817~\cite{Abbott:2018exr} event for the low- and high-mass posteriors. Finally, the inner (outer) gray region corresponds to the mass and radius constraint at 68.2\% (95.4\%) obtained for PSR J0030+0451 by the group analyzing NICER X-ray data~\cite{Miller:2021qha}.}
    \label{fig:m_r_band}
\end{figure*}

In this section, we discuss the influence of $\Delta$ matter on the EoS in the parity doublet model in the isospin-symmetric matter, as well as under the NS conditions of $\beta$-equilibrium and charge neutrality at $T=0$. 

In Fig.~\ref{fig:sigma_600}, we show the baryon chemical potential dependence of the expectation value of the $\sigma$ mean field for the isospin-symmetric matter. To illustrate the effects of the onset of $\Delta$ matter on the chiral structure, we fix $m_0^N=600~$MeV and study the dependence on the parameter $m_0^\Delta$. In all cases shown in the figure, the results exhibit similar behavior in the vicinity of the liquid-gas phase transition at small values of the baryon chemical potential. Thus, the appearance of $\Delta$ matter does not spoil the properties of the nuclear ground state. In the case of the purely nucleonic EoS, the chiral transition is a smooth crossover. The transition happens at $\mu_B\simeq 1400~$MeV and corresponds to a sudden drop of $\sigma$ to about zero, causing the parity-doublet nucleons ($N$, $N^\star$) to become nearly equally populated (cf.~Eq.~\eqref{eq:doublet_mass}). In general, the appearance of $\Delta$ matter is seen as a change in the stiffness of the EoS above the saturation density; thus, deviations from the purely nucleonic EoS can be attributed to the onset of $\Delta$ matter. For the case $m_0^\Delta=650~$MeV, $\sigma$ deviates smoothly from the purely nucleonic result around $\mu_B=1100~$MeV, which corresponds to the onset of positive-parity $\Delta$ matter. After this point, $\sigma$ changes very slowly, and the population of chiral partners is seen only mildly around $\mu_B=1400~$MeV and $\mu_B=1900~$MeV, for $N^\star$ and $\Delta^\star$, respectively. For the case $m_0^\Delta=600~$MeV, the swift decrease of $\sigma$ at $\mu_B=950$~MeV corresponds to the onset of positive-parity $\Delta$ through a first-order phase transition. The chiral symmetry gets restored around $\mu_B=1500~$MeV where the chiral partners of the nucleon and $\Delta$ appear almost simultaneously. We note that for smaller values of $m_0^\Delta$, $\Delta$ enters the matter at smaller baryon chemical potentials before $N^\star$ is populated in the purely nucleonic EoS. In both cases, it is associated with a drop of $\sigma$. At the same time, the onset of $\Delta^\star$, and thus the full restoration of chiral symmetry, is shifted to higher baryon chemical potentials.

In Fig.~\ref{fig:p_rho_600}, we plot the thermodynamic pressure for the same set of EoSs. In general, the chiral structure resembles the stiffness of the EoS. The softening of the pressure in the purely nucleonic EoS around $4~n_0$ corresponds to the chiral symmetry restoration and the onset of $N^\star$. Similar to the $\sigma$ expectation value, the onset of $\Delta$ matter is signaled by deviations from the template nucleonic EoS. For $m_0^\Delta=650~$MeV the onset is smooth and only softens the thermodynamic pressure. However, for $m_0^\Delta=600~$MeV, the onset of $\Delta$ through the first-order transition is pronounced in the density jump of the order of $0.5~n_0$. In general, smaller values of $m_0^\Delta$ result in the earlier onset of $\Delta$ matter. However, the strength of the transition depends on the choice of $m_0^\Delta$. Namely, higher values yield weaker first-order transition, which turns into a second-order and eventually becomes a smooth crossover, defined as a peak in $\partial \sigma / \partial \mu_B$. When $m_0^\Delta \rightarrow \infty$, the onset of $\Delta$ matter is shifted to higher densities and the EoS converges to the purely nucleonic case. We note that the structure discussed for $m_0^N=600~$MeV remains qualitatively the same for other values of the parameter.

\begin{figure}[t!]
  \centering
  \includegraphics[width=\linewidth]{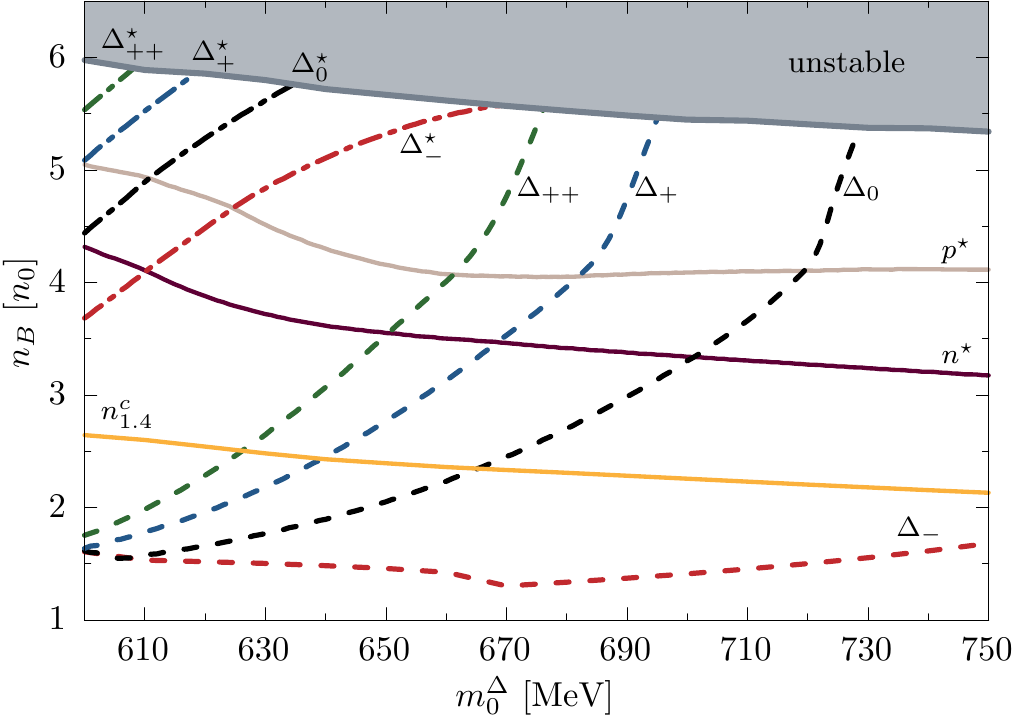}
  \caption{Onset densities of particle species for $m_0^N=600~$MeV as a function of $m_0^\Delta$. The gray-shaded regions indicate densities beyond the maximum mass configurations. The orange, solid line marks the central density of $1.4~M_\odot$ neutron stars. We note that the onset densities of the positive-parity nucleons, proton and neutron, are not shown in the figure.}
  \label{fig:onset_delta}
\end{figure}

In the left panel of Fig.~\ref{fig:pressure}, shown are the EoSs for selected values of $m_0^N$. To illustrate the effect of $\Delta$ matter on the EoS at intermediate densities, we show results obtained for purely nucleonic EoS (dashed line) together with the case $m_0^\Delta = m_0^N$ (solid line). The regions bounded by the two results correspond to the range spanned by solutions with $m_0^N < m_0^\Delta$ in each case. In the left panel of the figure, we show the isospin-symmetric matter EoSs. The orange-shaded region shows the proton flow constraint~\cite{Danielewicz:2002pu}. In general, the low-density behavior in each case is similar, until the deviations from the purely nucleonic EoSs are induced by the onset of $\Delta$ matter. The softening due to the onset of $\Delta$ resonance yields better agreement with the proton flow constraint. The EoSs with $\Delta$ converge back to the purely nucleonic EoS at high densities. 

In the right panel of Fig.~\ref{fig:pressure}, we show the corresponding EoSs under the neutron-star conditions. The grey- and orange-shaded envelopes show the constraints derived in~\cite{Abbott:2018exr} and~\cite{Annala:2019puf}, respectively. Similar to the isospin-symmetric case, the low-density behavior is comparable. For $m_0^N=550~$MeV, the EoSs with $m_0^\Delta \approx m_0^N$ result in an appearance of $\Delta_-$ via a strong first-order phase transition and underestimate the constraint at low densities. However, the phase transition is followed by a subsequent stiffening as compared to the purely nucleonic case and the EoS reaches the constraint at higher densities. It suggests that the onset of $\Delta$ matter softens the EoS and stiffens it at larger densities. This also resembles a higher speed of sound at intermediate densities. We note that this effect is more readily pronounced for smaller values of $m_0^\Delta$. For other parametrizations shown in the figure, the EoSs fall into the region derived by the constraint. We note that in the study of cold and dense QCD, commonly used are separate effective models for the hadronic and quark matter phases (two-phase approaches) with a priori assumed first-order phase transition, typically associated with simultaneous chiral and deconfinement transitions~\cite{Alvarez-Castillo:2016wqj}. As recently demonstrated in~\cite{Marczenko:2021uaj}, such a strong phase transition with large latent heat can occur also within hadronic matter due to the onset of $\Delta$ matter being subject to chiral symmetry restoration.

\begin{figure}[t!]
  \centering
  \includegraphics[width=\linewidth]{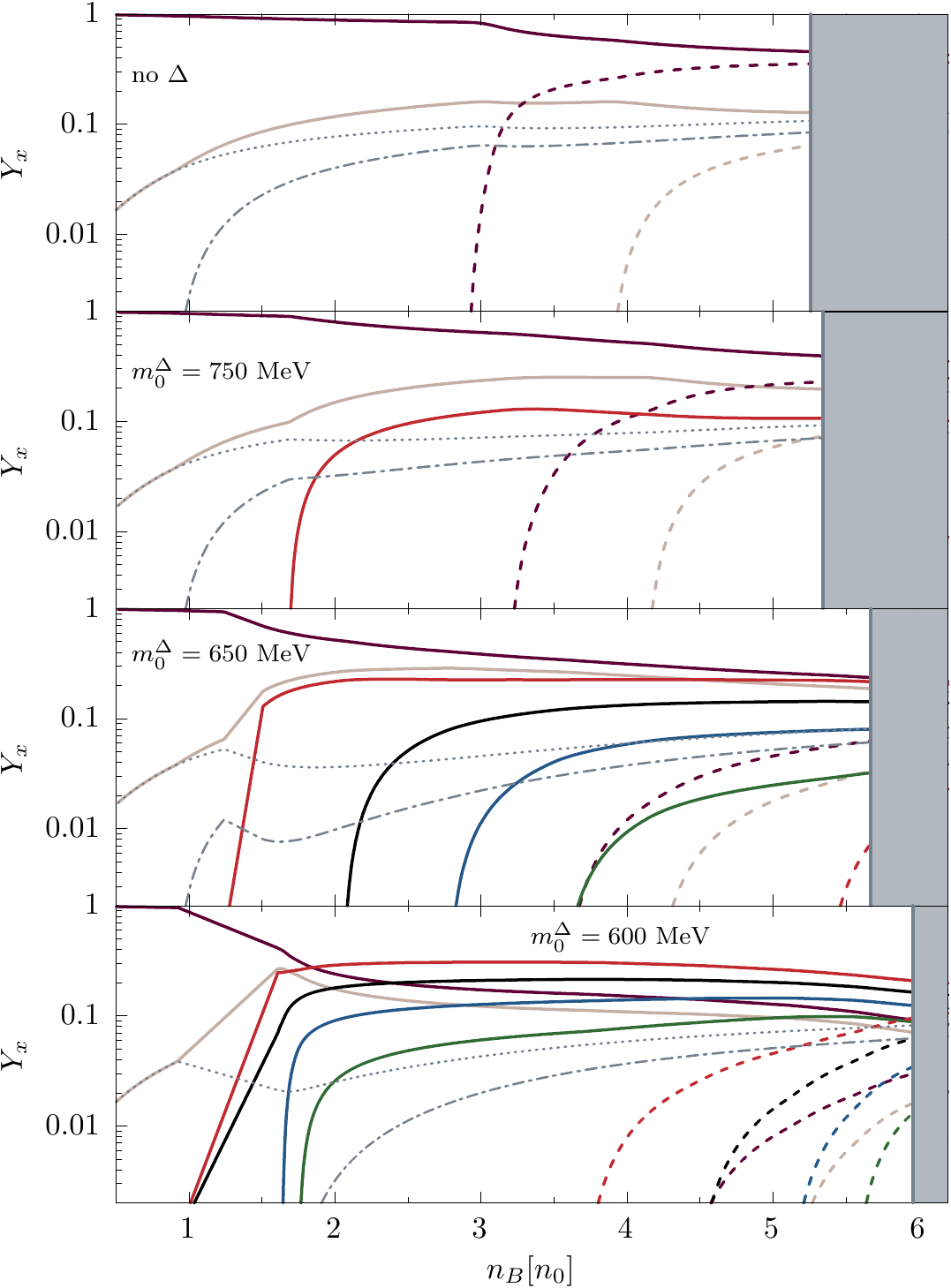}
  \caption{Particle fractions as a function of the net-baryon density in the units of saturation density for $m_0^N=600~$MeV and different values of $m_0^\Delta$. The top panel shows the results for the parity double model without $\Delta$ matter. The gray-shaded regions indicate densities beyond the maximum mass configurations. The solid (dashed) lines show the positive-parity (negative-parity) particles. The color coding is the same as in Fig.~\ref{fig:onset_delta}. The electrons and muons are shown as gray, dotted and dashed, lines, respectively.}
  \label{fig:composition}
\end{figure}

\section{Properties of neutron stars}
\label{sec:neutron_stars}

\subsection{TOV solutions}
\label{sec:tov}

\begin{figure}[t!]
  \centering
  \includegraphics[width=\linewidth]{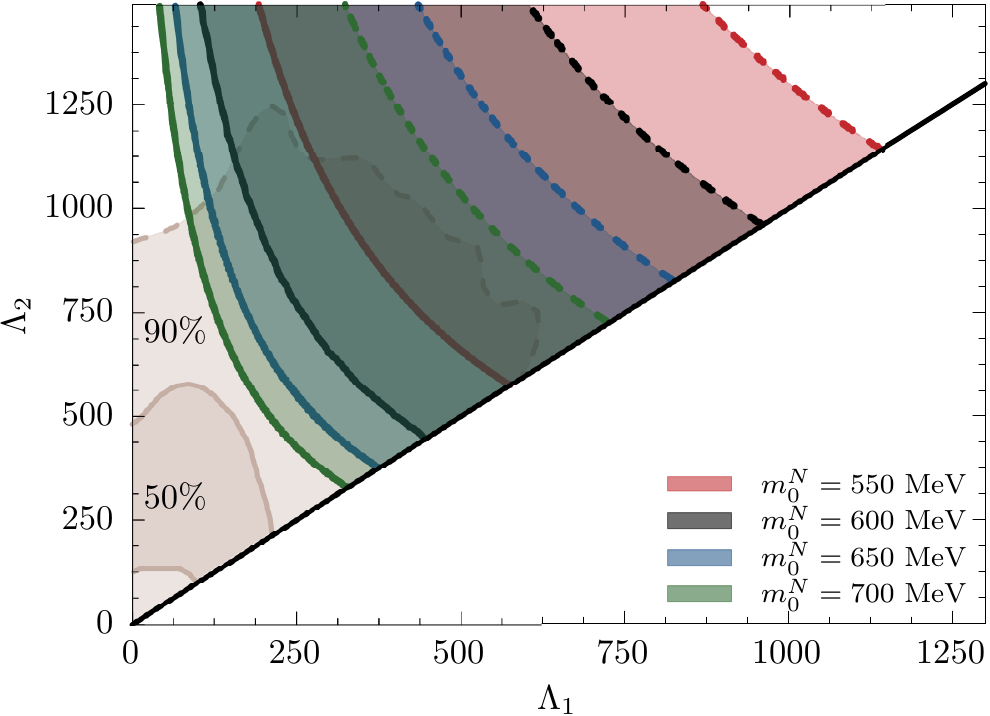}
  \caption{The relation between tidal deformabilities $\Lambda_1$ and $\Lambda_2$ of two compact stars that merged in the GW170817 event~\cite{Abbott:2018exr}. For comparison, also shown are the 50\% and 90\% fidelity regions from the analysis of the GW signal by the LIGO-VIRGO Collaboration~\cite{Abbott:2018exr}. Shown is only the physical region where $\Lambda_1 < \Lambda_2$.}
  \label{fig:L1_L2_band}
\end{figure}

In this section, we explore the impact of the emergence of the $\Delta$ matter at supersaturation densities on the structure of neutron stars. The composition of neutron-star matter requires $\beta$-equilibrium with leptons ($l$): electrons ($e$) and muons ($\mu$), included as free relativistic particles, as well as the charge neutrality condition. 

The EoS of dense matter plays a crucial role in determination of the structure of neutron stars. Its microscopic details are linked to the specific values of mass and radius, and therefore compactness. We use the EoSs introduced in the previous section to solve the general-relativistic Tolman-Oppenheimer-Volkoff (TOV) equations~\cite{Tolman:1939jz, Oppenheimer:1939ne} for spherically symmetric objects, 
\begin{subequations}\label{eq:TOV_eqs}
\begin{align}
   \frac{\dd P(r)}{\dd r} &= -\frac{\left(\epsilon(r) + P(r)\right)\left(M(r) + 4\pi r^3 P(r)\right)}{r \left(r-2M(r)\right)} \textrm,\\
   \frac{\dd M(r)}{\dd r} &= 4\pi r^2 \epsilon(r)\textrm,
\end{align}
\end{subequations}
with the boundary conditions \mbox{$P(r=R) = 0$}, \mbox{$M = M(r=R)$}, where $R$ and $M$ are the radius and the mass of a neutron star, respectively. Once the initial conditions are specified based on a given equation of state, namely the central pressure $P_c$ and the central energy density $\epsilon_c$, the internal profile of a neutron star can be calculated.

In general, there is one-to-one correspondence between the EoS and the \mbox{mass-radius} relation calculated from Eqs.~\eqref{eq:TOV_eqs}. In the left panel of Fig.~\ref{fig:m_r_band}, we show the relationship of mass versus central net-baryon density, for the calculated sequences of neutron stars for $m_0^N=550~$MeV, $m_0^N=600~$MeV, $m_0^N=650~$MeV, $m_0^N=700~$MeV, together with the state-of-the-art constraint on the maximum mass for the pulsar PSR J0740+6620, $M=2.08\pm0.07~M_\odot$~\cite{Fonseca:2021wxt}. Shown are sequences for $m_0^\Delta=m_0^N$ (solid line) and pure nucleonic sequence (dashed line). The color-shaded areas show the region spanned by solutions obtained with $m_0^\Delta > m_0^N$. The relations are plotted up to the maximally stable solutions. We point out that, similarly to the EoS, the onset of positive-parity $\Delta$ leads to a softening of the mass-radius sequence so that it is accompanied by a strong increase of the central density. This is seen for $m_0^N = m_0^\Delta$ with plateaux at small NS masses. We note that in general smaller values of $m_0^N$ yield larger values of the maximum mass, owing to the stiffness of the EoS. 

In the right panel of Fig.~\ref{fig:m_r_band}, we show the relationship of mass and radius. Also shown are the state-of-the-art constraints: the high precision mass-radius analysis of the massive pulsar PSR~J0740+6620 by the NICER collaboration~\cite{Riley:2021pdl, Miller:2021qha}, constraint from the recent GW170817 event~\cite{Abbott:2018exr}, and the constraint obtained for PSR J0030+0451 by the group analyzing X-ray data~\cite{Miller:2019cac}. The appearance of $\Delta$ matter affects the NS structure which is reflected in the mass-radius relations. As discussed in Sec.~\ref{sec:eos}, the softening of the EoS due to the appearance of $\Delta$ is followed by a subsequent stiffening of the EoS. A substantial reduction of the radii is observed. For instance, the radii of the $1.4~M_\odot$ NS obtained in the purely nucleonic EoS reduce roughly up to $1~$km for $m_0^N=700~$MeV, to almost $2~$km for $m_0^N=550~$MeV. On the other hand, the decrease of the star's maximum mass is seen only mildly. This, in turn, is a consequence of the subsequent stiffening of the EoS at higher densities, which allows for massive stars to sustain from the gravitational collapse. In Ref.~\cite{Li:2018qaw}, it was shown that the earlier onset of $\Delta$ yields a larger maximum mass. We note that this stays in contrast to our work, where we find a mild decrease of the maximum mass. This is due to additional softening of the EoS at high densities that is provided by the onset of the negative-parity $N^\star$ and $\Delta^\star$ due to the chiral symmetry restoration. This highlights the importance of chiral symmetry restoration in the EoS and the properties of the NS matter.

In Fig.~\ref{fig:onset_delta} we study the onset densities of the particle species. As an example, we show the results for a fixed value of $m_0^N=600~$MeV. We note that the onset densities of the positive-parity nucleons, proton, and neutron, are not shown in the figure. Their negative-parity counterparts, $n^\star$ and $p^\star$, are always present for any choice of $m_0^\Delta$. They enter the matter composition roughly between $3-5n_0$. Their appearance is always sequential due to differences in their effective chemical potentials. Conversely, for small values of $m_0^\Delta$, the positive-parity $\Delta$ resonances appear almost simultaneously through a first-order transition, which triggers a sufficiently big density jump. At higher values of the parameter, the differences in the onset density become more readily exposed. The onset density of $\Delta_-$ changes only mildly within the shown range of the parameter, while the densities for others increase and eventually the $\Delta_{++}$, $\Delta_+$, and $\Delta_0$ are being populated in the gravitationally unstable branch of the EoS. We find similar behavior for the negative-parity $\Delta^\star$'s. They appear sequentially and no $\Delta^\star$'s are present for $m_0^\Delta > 670~$MeV. We stress that similarly to matter constituents in symmetric matter, the appearance of the chiral partners, reflects the partial chiral symmetry restoration. For $m_0^N=600~$MeV the $\Lambda_{1.4}<580$ constraint is met for $m_0^\Delta<650~$MeV (see Ref.~\cite{Marczenko:2021uaj}). Therefore, the appearance of $\Delta$ matter is essential for softening the EoS at intermediate densities. Although $\Delta^\star$'s are not populated at central densities of $1.4~M_\odot$ NSs, the compliance with the deformability constraint implies their onset at higher, gravitationally stable, densities,  and therefore, their presence in the cores of high-mass neutron stars. This is a direct consequence of the chiral symmetry restoration. Namely, the onset of $\Delta$ matter drastically decreases the value of the $\sigma$ mean-field,  and consequently, decreases the mass of $\Delta^\star$'s towards the asymptotic value of $m_0^\Delta$, allowing for their earlier onset. For sufficiently high $m_0^\Delta$, $\Delta$ matter is not populated in the gravitationally stable part of the mass-radius sequence. The shift of the threshold of the onset density is qualitatively similar as in the case of increasing the strength of repulsion between $\Delta$'s, which may push their onset out of the gravitationally stable branch of the stellar sequence. Thus, the corresponding EoSs would be equivalent to the purely nucleonic EoS in this range. Lastly, we note that the maximum-mass central density and the central density of a $1.4~M_\odot$ NS decrease only mildly by approximately $0.5~n_0$ as the value of $m_0^\Delta$ increases.

\begin{figure}[t!]\centering
  \includegraphics[width=\linewidth]{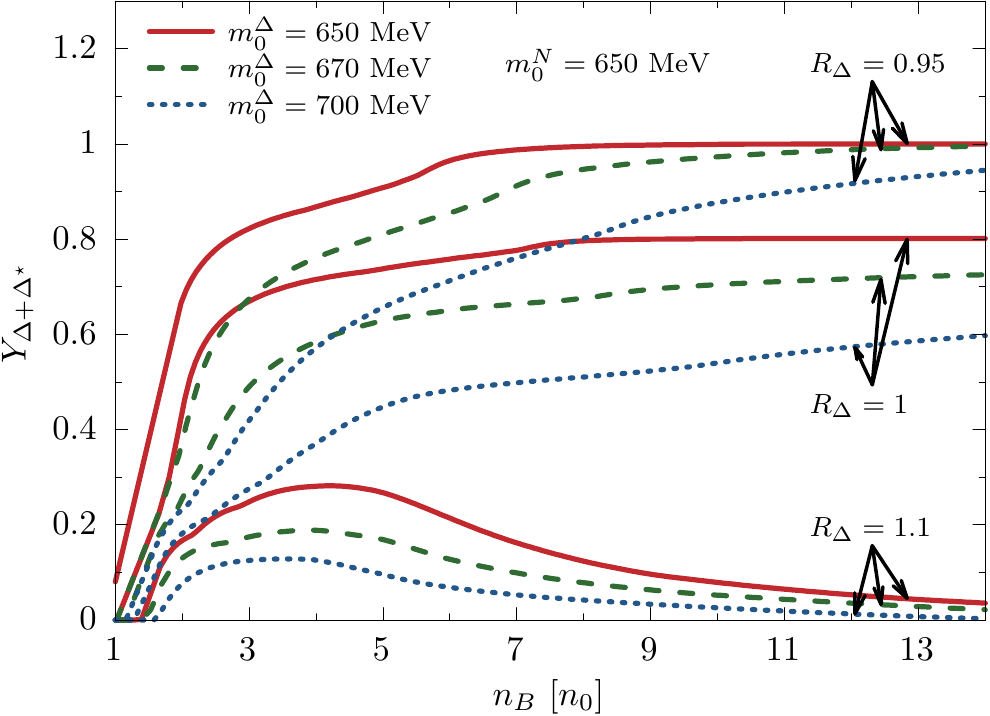}
  \caption{$\Delta$ matter fraction under the NS conditions of $\beta$ equilibrium and charge neutrality as a function of the net-baryon density in the units of saturation density.}
  \label{fig:delta_rep_fraction}
\end{figure}

In Fig.~\ref{fig:composition}, we show the neutron-star matter composition for $m_0^N=600~$MeV, in the case without $\Delta$ (first panel), $m_0^\Delta=750~$MeV (second panel), $m_0^\Delta=650~$MeV (third panel), $m_0^\Delta=600~$MeV (fourth panel). The gray-shaded regions mark the central densities that are gravitationally unstable. Because of charge neutrality, the inclusion of $\Delta$ and $\Delta^\star$ influences the composition of neutron stars. We observe that the onset of $\Delta^-$, and consequently other $\Delta$ states, is shifted to higher values of $m_0^\Delta$. Negatively charged $\Delta_-$ and $\Delta_-^\star$ partially replace the role of muons in compensating the charge of positively charged baryons. For instance, the proton fraction drastically increases upon the onset of $\Delta_-$. This may have consequences for the thermal evolution of NSs and different cooling mechanisms, which are sensitive to the proton fraction~\cite{Klahn:2006ir}. The direct Urca processes that include the $\Delta$ resonance may be essential for rapid cooling of NSs as they can become operative without the presence of exotic states nor the large proton concentration ($11-14\%$), which is required for the canonical direct Urca process~\cite{Prakash:1992zng}. Moreover, the threshold proton concentration for the direct Urca process reduces by about 3\% when the chiral symmetry is restored, i.e., nucleons and their chiral partners are equally populated~\cite{Marczenko:2018jui}.

\begin{figure}[t!]\centering
  \includegraphics[width=\linewidth]{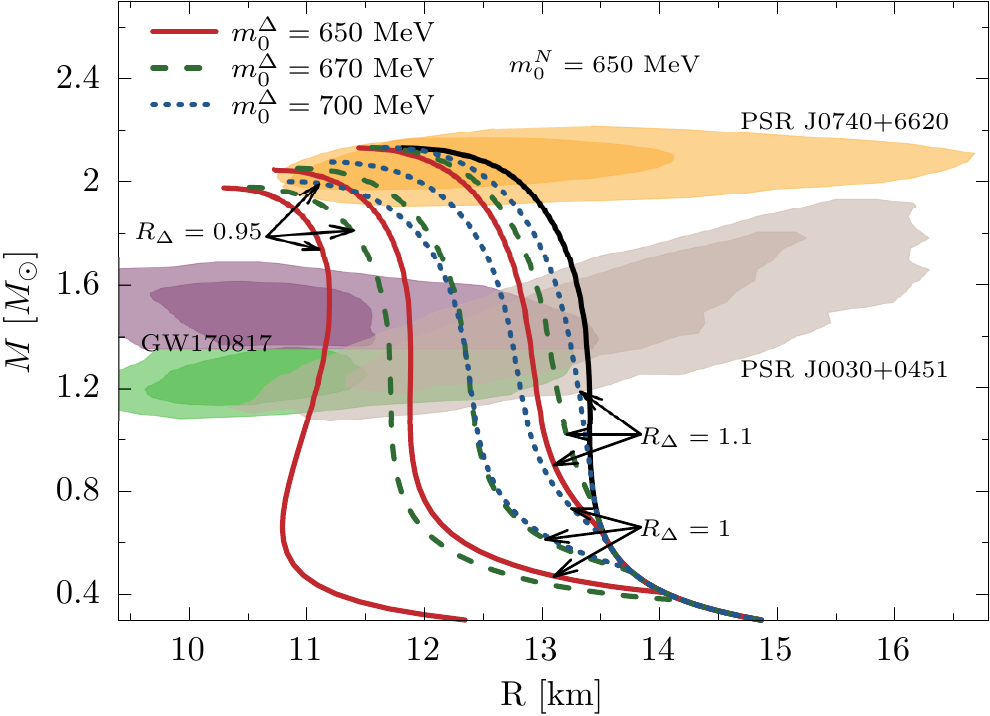}
  \caption{Mass-radius sequences for different values of the parameter $R_\Delta$ for $m_0^N=650~$MeV. Black, solid line shows the mass-radius sequences obtained for purely nucleonic EoS.}
  \label{fig:delta_rep_m_r}
\end{figure}

\subsection{Tidal deformability}
\label{sec:tidal}

\begin{figure*}[t!]\centering
  \includegraphics[width=.8\linewidth]{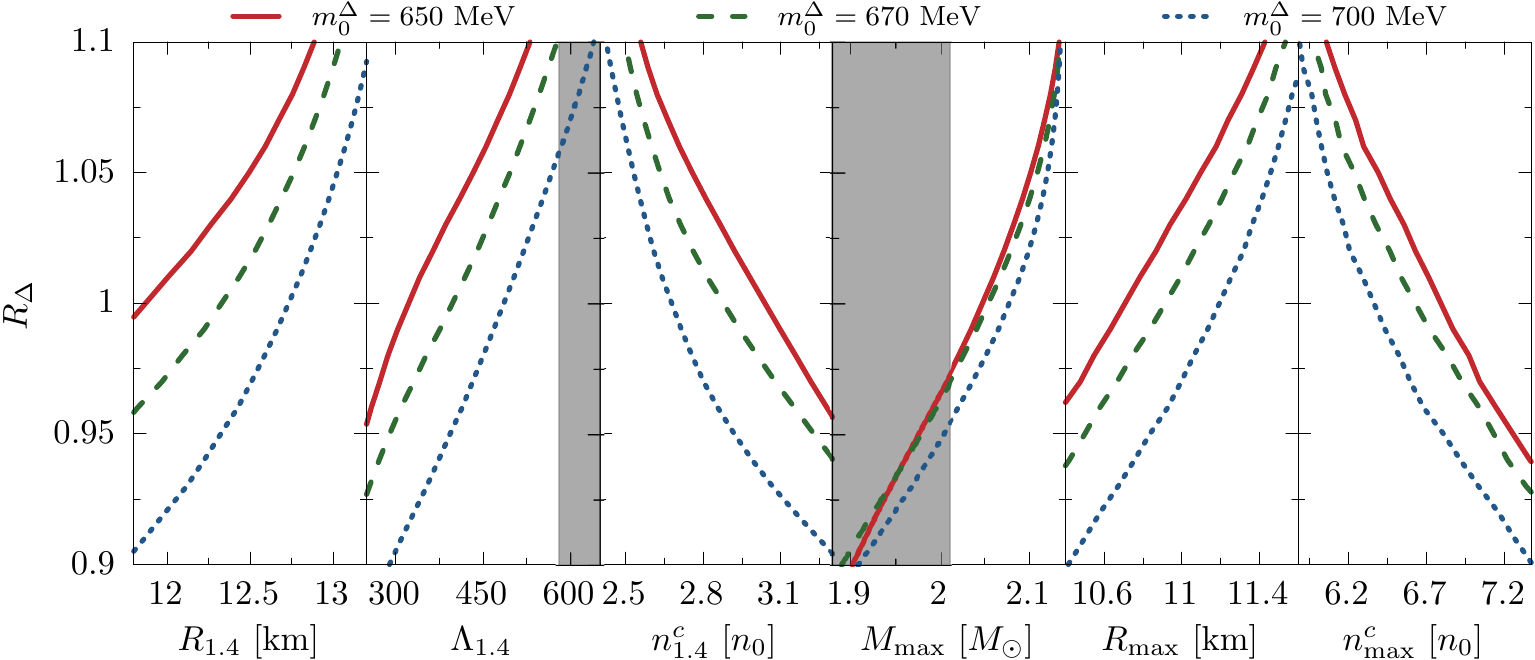}
  \caption{Various NS properties as functions of the parameter $R_\Delta$ for $m_0^N=650~$MeV and different values of $m_0^\Delta$. Gray-shaded areas show regions excluded by the astrophysical constraints (see text for details).}
  \label{fig:R_delta_NS_params}
\end{figure*}

The dimensionless tidal deformability parameter $\Lambda$ can be computed through its relation to the Love number $k_2$~\cite{Hinderer:2007mb,Damour:2009vw,Binnington:2009bb,Yagi:2013awa,Hinderer:2009ca},
\begin{equation}
   \Lambda = \frac{2}{3} k_2 C^{-5} \textrm,
\end{equation}
where $C = M/R$ is the star compactness parameter, with $M$ and $R$ being the total mass and radius of a star. The Love number $k_2$ reads
\begin{equation}
\begin{split}
   k_2 &= \frac{8C^5}{5} \left(1-2C\right)^2 \left[ 2+2C(y-1)-y \right] \times \\
      \times & \Big(2C\left[6- 3y + 3C(5y-8) \right]\\
      +      & 4C^3[13-11y+C(3y-2)+2C^2(1+y)] \\
      +      & 3(1-2C)^2[2-y+2C(y-1)\ln{(1-2C)}]\Big)^{-1} \textrm,
\end{split}
\end{equation}
where $y = R\beta(R)/H(R)$. The functions $H(r)$ and $\beta(r)$ are given by the following set of differential equations:
\begin{eqnarray}
\frac{\dd \beta}{\dd r}&=&2 \left(1 - 2\frac{M(r)}{r}\right)^{-1} \nonumber\\
&& H\left\{-2\pi \left[5\epsilon(r)+9 P(r)+\frac{\dd \epsilon}{\dd P}(\epsilon(r)+P(r))\right]\phantom{\frac{3}{r^2}} \right. \nonumber\\
&& \left.+\frac{3}{r^2}+2\left(1 - 2\frac{M(r)}{r}\right)^{-1} \left(\frac{M(r)}{r^2}+4\pi r P(r)\right)^2\right\}\nonumber\\
&&+\frac{2\beta}{r}\left(1 - 2\frac{M(r)}{r}\right)^{-1}\nonumber\\
&&  \left\{\frac{M(r)}{r}+2\pi r^2 (\epsilon(r)-P(r)) - 1\right\}~\textrm,\\
\frac{\dd H}{\dd r}&=& \beta \textrm.
\end{eqnarray}
The above equations have to be solved along with the TOV equations~(\ref{eq:TOV_eqs}). The initial conditions are \mbox{$H(r\rightarrow 0) = c_0 r^2$} and \mbox{$\beta(r\rightarrow 0) =2c_0 r$}, where $c_0$ is a constant, which is irrelevant in the expression for the Love number $k_2$.

In Fig.~\ref{fig:L1_L2_band}, we plot the tidal deformability parameters $\Lambda_1$ vs $\Lambda_2$ of the high- and low-mass members of the binary merger together with the 50\% and 90\% fidelity regions obtained by the LVC analysis of the GW170817 event~\cite{Abbott:2018exr}. We note that the tidal deformability parameter requires sufficiently soft EoS up to a few times the saturation density. This is seen in the figure, where the smallest tidal deformability and thus the best agreement with the constraint is obtained for smallest values of $m_0^\Delta$, which corresponds to softest EoSs at low density. On the other hand, the $2~M_\odot$ requires a sufficiently stiff equation of state at higher densities. Inversely to the tidal deformability, the most massive stars are obtained for the stiffest EoSs, namely for purely nucleonic ones. Therefore, the two constraints are exclusive, which allows for their precise determination. In Ref.~\cite{Marczenko:2021uaj}, it was shown that the chirally invariant masses can be roughly estimated to lie in the range from 550 to 680~MeV. We remark that these values are at tension with the results obtained in the LQCD simulations~\cite{Aarts:2017rrl} at vanishing chemical potential and finite temperature where within the errors the masses of the ground state nucleon and $\Delta$ do not deviate from their vacuum masses. This might suggest additional medium dependence of the chirally invariant masses,
whereas the LQCD simulations were performed with large pion mass, $m_\pi = 400$ MeV.
We also note that the interplay between the $2~M_\odot$ and the tidal deformability constraints can be further used to fix the allowed range of external model parameters. This would be of particular use in a class of effective models in which the low- and high-density regimes are not treated independently but rather combined in a consistent unified framework.

\begin{table*}[t!]\centering\begin{tabular}{|c|c||c|c|c|c||c|c|c|c|}
  \hline
  $R_\Delta$ & $m_0^\Delta~$[MeV] & $R_{1.4}~$[km] & $n^c_{1.4}~[n_0]$ & $\Lambda_{1.4}$ & $Y^{1.4}_{\Delta+ \Delta^\star}$ & $R_{\rm max}~$[km] &$n^c_{\rm max}~[n_0]$ & $M_{\rm max}~[M_\odot]$ & $Y^{\rm max}_{\Delta + \Delta^\star}$ \\ \hline\hline
    \multirow{3}{*}{0.95} & 650         & 11.2 & 3.34 & 245 & 0.84 & 10.3 & 7.25 & 1.97 & 0.99 \\ \cline{2-9}
                          & 670         & 11.7 & 3.23 & 290 & 0.70 & 10.5 & 7.13 & 1.98 & 0.92 \\ \cline{2-9}
                          & 700         & 12.3 & 2.92 & 393 & 0.40 & 10.8 & 6.81 & 2.00 & 0.75 \\ \hline\hline
    \multirow{3}{*}{1.00} & 650         & 11.9 & 3.04 & 322 & 0.67 & 10.7 & 6.78 & 2.04 & 0.77 \\ \cline{2-9}
                          & 670         & 12.3 & 2.88 & 397 & 0.47 & 10.9 & 6.64 & 2.05 & 0.66 \\ \cline{2-9}
                          & 700         & 12.7 & 2.67 & 486 & 0.25 & 11.2 & 6.36 & 2.08 & 0.49 \\ \hline\hline
    \multirow{3}{*}{1.10} & 650         & 12.9 & 2.56 & 531 & 0.22 & 11.4 & 6.06 & 2.13 & 0.21 \\ \cline{2-9}
                          & 670         & 13.1 & 2.50 & 577 & 0.16 & 11.5 & 5.97 & 2.13 & 0.13 \\ \cline{2-9}
                          & 700         & 13.2 & 2.43 & 639 & 0.11 & 11.6 & 5.89 & 2.13 & 0.07 \\ \hline\hline
    \multicolumn{2}{|c||}{no $\Delta$}  & 13.4 & 2.30 & 701 & 0.00 & 11.8 & 5.76 & 2.13 & 0.00 \\ \hline
  \end{tabular}
  \caption{Dependence of the NS properties on the parameter $R_\Delta$ for $m_0^N=650~$MeV and different values of $m_0^\Delta$. The bottom row contains values obtained in purely nucleonic EoS.}
  \label{tab:delta_rep_m_r}
\end{table*}

\subsection{\texorpdfstring{$R_\Delta$}{} dependence}
\label{sec:repulsion_delta}

In this section, we explore the influence of the $\Delta$ couplings to vector mesons on the properties of dense matter. We start by considering the isospin-symmetric case, i.e., $\mu_Q=0$, which also implies that $\rho=0$. Firstly, we analyze the asymptotic matter composition. Due to chiral symmetry restoration at high density, baryon masses converge to constant values $m_0^x$~(cf. Eq~\eqref{eq:doublet_mass}). Because $m_0^x \ll \mu_B$, their contribution can be neglected at large values of chemical potential, thus
\begin{equation}
n_B^x \approx \frac{d_x}{6\pi^2} \mu_x^3\textrm,
\end{equation}
where $d_x$ and $\mu_x$ are the degeneracy factor and effective chemical potential of the $x$'th species, respectively. The effective chemical potential of the nucleon is \mbox{$\mu_N = \mu_B - g_\omega^N\omega$}. The effective chemical potential for $\Delta$ can be written as follows
\begin{equation}
\mu_\Delta = \mu_B - g_\omega^\Delta\omega = \mu_N - \left(R_\Delta-1\right)g_\omega^N\omega\textrm,
\end{equation}
thus, their difference
\begin{equation}\label{eq:chem_diff}
\mu_N - \mu_\Delta = \left(R_\Delta-1\right)g_\omega^N\omega\textrm, 
\end{equation}
is directly proportional to the value of $\omega$ mean field. In general, the value of $\omega$ is expected to increase with density (cf.~Eq.~\eqref{eq:gap_eqs}). Thus, the difference given in Eq.~\eqref{eq:chem_diff} increases with density as well. For $R_\Delta=1$, the chemical potentials are equal, i.e., $\mu_N = \mu_\Delta$. The densities of nucleons and $\Delta$ approach the same value and the matter composition can be determined solely based on their degeneracy factors, i.e., $Y_x = d_x/\sum_i d_i$. In this case, the fraction of $\Delta$ matter amounts to $Y_{\Delta + \Delta^\star}=0.8$. For $R_\Delta < 1$, $\mu_\Delta > \mu_N$ and their difference grows with density. The partial density $Y_{\Delta +\Delta^\star}$ will asymptotically approach unity. Conversely, for $R_\Delta>1$, $\mu_\Delta < \mu_N$, thus the system will be dominated by nucleons, and $Y_{\Delta + \Delta^\star}$ will vanish at high density. 

We note that the above analysis can be also performed for matter under NS conditions of $\beta$ equilibrium and charge neutrality. This case is depicted in Fig.~\ref{fig:delta_rep_fraction}, where shown are the results obtained for three values of the parameter $R_\Delta=0.95,~1,~1.10$ (see Eq.~\eqref{eq:r_delta}) for $m_0^N=650$~MeV and $m_0^\Delta=650,~670,~700~$MeV. The results converge to the appropriate asymptotic values for different values of $R_\Delta$. 

Densities in the cores of neutron stars can reach the values of the order of $5-8~n_0$ (cf.~Fig.~\ref{fig:m_r_band}). Already under such conditions, qualitative differences in the abundance of $\Delta$ matter are vividly seen in Fig.~\ref{fig:delta_rep_fraction}, even for small deviations of $R_\Delta$ from unity. For instance, in the considered examples, for a canonical $1.4~M_\odot$ NS $\Delta$ matter concentration can be up to four times smaller for $R_\Delta=1.10$ when compared to $R_\Delta=0.95$. Interestingly, we find that for $R_\Delta>1$ the fraction of $\Delta$ matter exhibits a maximum at few times $n_0$, owing to the asymptotic behavior, which requires that $\Delta$ matter should disappear from the system at high densities. Similarly for $R_\Delta<1$, the matter at high densities is composed solely of $\Delta$s, and the nucleons are not present. We note that different studies suggested various ranges of the repulsive forces (see, e.g,~\cite{Maslov:2015msa, Drago:2014oja}). The differences in the matter composition due to the repulsive interactions are also seen in the bulk NS properties. In Fig.~\ref{fig:delta_rep_m_r}, shown are the corresponding mass-radius sequences. As expected, more compact solutions are obtained for smaller values of $m_0^\Delta$ and $R_\Delta$. Contrary to the previous finding that the decrease of $m_0^\Delta$ changes the maximum mass only mildly, smaller values of $R_\Delta$ yield noticeably smaller maximum masses. In Fig.~\ref{fig:R_delta_NS_params}, we plot various NS properties as functions of $R_\Delta$. Besides the decrease in radius of the canonical $1.4~M_\odot$ NS and the maximum mass for smaller values of $R_\Delta$, we also observe a notable decrease in the tidal deformability parameter $\Lambda_{1.4}$.  Therefore, as far as the repulsive interactions are concerned, the matter composition inside the cores of $2~M_\odot$ NSs may be very  different upon small variations in the strength of the repulsive interactions among $\Delta$ isobars. Thus, determining the NS matter composition has to be treated with particular care. For completeness, the NS properties are listed in Table~\ref{tab:delta_rep_m_r}.

\section{Summary}
\label{sec:summary}

We have explored the influence of the formation of $\Delta$ resonance in dense matter on the bulk properties of neutron stars (NSs). To this end, we employed the parity doublet model for nucleons and $\Delta$'s, which accounts for the self-consistent treatment of the chiral symmetry
breaking and its restoration in the mesonic and baryonic sectors. We analyzed the equation of state (EoS) under the NS conditions of $\beta$-equilibrium and charge neutrality.

We have shown how modern astrophysical constraints on the maximum mass~\cite{Fonseca:2021wxt}, the tidal deformability from the binary merger GW170817~\cite{Abbott:2018exr}, and recent simultaneous mass-radius constraints from the NICER experiment~\cite{Miller:2021qha, Miller:2019cac, Riley:2019yda, Riley:2021pdl} allow us for a consistent determination of the EoS of dense matter. We have demonstrated that the purely nucleonic EoSs obtained in the parity doublet model are too stiff at intermediate densities and therefore they are ruled out by the recently revised tidal deformability constraint. This is caused by the apparent correlation between the properties of matter at saturation and the compactness at a few times the saturation density.

We find that the early appearance of $\Delta(1232)$ resonance softens the EoS at intermediate densities. As a result, the radius of canonical $1.4~M_\odot$ reduces substantially providing better agreement with the tidal deformability constraint from the GW170817 event. We observe a subsequent stiffening of the EoS at higher densities. As a consequence, we also find, in contrast to previous studies~\cite{Li:2018qaw}, that the maximum mass decreases compared to purely nucleonic results, and stays in agreement with observational astrophysical data. This is traced back to the restoration of chiral symmetry, which further softens the EoS at higher densities and results in the onset of $N^\star$ and  $\Delta^\star$ resonances in the high-mass part of the NS sequence.

We emphasize that an abrupt change in mass-radius profile in the high-mass part of the sequence  can be in general due to phase transition in nuclear matter. However, as discussed above and also shown in~\cite{Marczenko:2021uaj, Somasundaram:2021clp}, such a change is not necessarily linked to a hadron-quark phase transition that implies the existence of quark matter in the cores of NSs, as suggested in~\cite{Alvarez-Castillo:2016wqj, Li:2021sxb, Christian:2021uhd}.

To address in more detail the role of baryonic resonances in the presence of chiral symmetry restoration, we have analyzed the repulsive $\Delta$-(vector meson) interactions and studied their consequences for the phenomenological description of NSs. We find that small repulsion yields more compact stellar sequences with reduced maximum mass, and that variations of the strength of these interactions lead to notable qualitative changes in the matter composition at a few times the saturation density. Thus, the determination of the structure of the  NS matter requires detailed knowledge of the intricate repulsive forces between individual hadronic constituents.

In the present study, we have assumed a hadronic EoS to conclude the composition of neutron star interiors and the phase structure of dense matter. A natural extension is to generalize the parity doublet model with $\Delta$ and include the quark degrees of freedom that may allow the existence of hybrid quark-hadron stars through a deconfinement phase transition while fulfilling the maximum mass constraint~\cite{Marczenko:2020jma}. Furthermore, it is interesting to verify if signatures of the chiral symmetry restoration related to a crossover or first-order phase transition in the hadronic phase have an observable imprint on the GW emission from NS mergers, and, thus, can be measurable in future GW detections.

Given the recent formulation of the three-flavor parity doubling~\cite{Steinheimer:2011ea, Sasaki:2017glk} and further lattice QCD studies~\cite{Aarts:2015mma, Aarts:2017rrl}, which indicate  that the parity doubling occurs also in the hyperon channels, it would be of great interest to extend this work and include contributions of hyperons. 

Furthermore, one anticipates that the $\Delta$ resonance may be essential for rapid cooling of NSs~\cite{Prakash:1992zng}. Thus, it is alluring to examine how the presence of the chiral partners of the nucleon and $\Delta$ affect the NS thermal evolution. Work in these directions is in progress.

\begin{acknowledgements}
This work is supported partly by the Polish National Science Centre (NCN) under OPUS Grant No. 2018/31/B/ST2/01663 (K.R. and C.S.), Preludium Grant No. 2017/27/N/ST2/01973 (M.M.), and the program Excellence Initiative–Research University of the University of Wroclaw of the Ministry of Education and Science (M.M). K.R. also acknowledges the support of the Polish Ministry of Science and Higher Education. M.M. acknowledges helpful discussions with David E. Alvarez-Castillo. We also acknowledge fruitful discussions with David Blaschke and Tobias Fischer.  
\end{acknowledgements}

\bibliographystyle{apsrev4-1}
\bibliography{biblio}

\end{document}